\documentclass[a4paper,11pt]{article}

\usepackage[utf8]{inputenc}
\usepackage[english]{babel}
\usepackage{amsmath}
\usepackage{amssymb}
\usepackage{physics}
\usepackage{units}
\usepackage{cite}                       
\usepackage{hyperref}
\usepackage{graphicx}
\usepackage[margin=2.8cm]{geometry}     
\usepackage{tikz}                       
\usepackage[width=0.8\textwidth]{caption}  
\usepackage[symbol]{footmisc}           
\usepackage{titlesec}                   

\titleformat{\paragraph}[display]{\it}{}{}{}

\title{Dynamical relaxation in 2HDM models}

\author{
Zygmunt Lalak\thanks{Zygmunt.Lalak@fuw.edu.pl} \;
Adam Markiewicz\thanks{Adam.Markiewicz@fuw.edu.pl}\vspace{0.5em}\\
\small\it Institute of Theoretical Physics, Faculty of Physics, University of Warsaw\\
\small\it ul. Pasteura 5, 02-093 Warsaw, Poland
}

\date{}

\begin{document}

\maketitle

\begin{abstract}
The dynamical relaxation provides an interesting solution to the hierarchy problem in face of the missing signatures of any new physics in recent experiments. Through a~dynamical process taking place in the inflationary phase of the universe it manages to achieve a~small electroweak scale without introducing new states observable in~current experiments. Appropriate approximation makes it possible to derive an explicit formula for the final vevs in the double-scanning scenario extended to a model with two Higgs doublets (2HDM). Analysis of the relaxation in 2HDM confirms, that in a~general case it is impossible to keep vevs of both scalars small, unless fine-tuning is present or additional symmetries are cast upon the Lagrangian. Within the slightly constrained variant of 2HDM, where odd powers of the fields' expectation values are not present (which can be easily enforced by requiring that the doublets have different gauge transformations or by imposing a global symmetry) it is shown  that the the difference between the vevs of two scalars tends to be proportional to the cutoff. The analysis of the relaxation in 2HDM indicates, that in a~general case the relaxation would be stopped by the first doublet that gains a vev, with the other one remaining vevless with a mass of the order of the cutoff. This happens to conform with the inert doublet model.
\end{abstract}

\section{Introduction}

It is well  known  that the large quantum corrections to the Higgs' mass present in Standard Model do not fit with its small observed value (and the related small value of the electroweak scale). If one wants to avoid a~fine-tuned cancellation an extension of the SM must be used. Traditional solutions like supersymmetry or extra dimensions predict new physics visible at energies close to the electroweak scale. However, prolonged absence of new effects in experiments forces one to look for models that maintain naturalness and produce a~hierarchy of scales at the same time. Lack of new physics in the recent LHC results makes the need for such models more pressing than ever~\cite{Graham2015,Espinosa2015,Hardy2015,Gupta2015,Choi2015,Ibanez2015,Kobayashi2016,Hook2016,Choi2016,Flacke2016,McAllister2016}.

A recent attempt at obtaining a~scale hierarchy with technically natural parameters is the dynamical relaxation. In this scenario the electroweak scale is selected through a~dynamical process guided by an interaction of the Higgs doublet with new scalar fields. As those fields are very weakly interacting, they can easily avoid detection. At the same time the mechanism allows for pushing the new physics scale as far as \(\unit[10^9]{GeV}\)~\cite{Espinosa2015}. Given multiple models that include more then one Higgs doublet (most notably supersymmetric extensions of the Standard Model) it is interesting to see how does the relaxation mechanism perform in those scenarios and to find out conditions which must be satisifed for it to work.

The first part of this note describes the computation of the precise value of the electroweak scale in two basic variants of the relaxation scenario. The first one is the original idea of the relaxation, that requires only one additional scalar field. Although  it was shown to be challenged by observational data, it still serves as an useful introduction and clearly explains the fundamental concept. The second model is an extension of the first, with one more field required, which leads to the so called double-scanner mechanism.

The cosmological evolution predicted by the double-scanning mechanism is explained step-by-step, with the prime result being an explicit formula for the final electroweak scale in terms of the parameters present in the Lagrangian.

The second part deals with possible extensions of the double-scanning to a~model with two Higgs doublets. A~general construction is presented and the final electroweak scale is calculated in a~special case that allows for explicit analytical solution. It is shown that naturally the double-scanning mechanism can keep at most one of the vevs small.

\section{Dynamical relaxation model}

The simplest relaxion based model, first presented in \cite{Graham2015}, supplements the Standard Model with axion-like field \(\phi\), termed relaxion in this context. Contrary to the usual QCD axion this field is required to have a~large, non-compact range. Additionally the model introduces a~soft symmetry-breaking coupling to Higgs. Together, this produces a~dynamical mechanism, which naturally results in a~hierarchy between the electroweak scale and the model's cutoff. The full potential relevant for the process is given by
\begin{equation}
	\label{eq:basic_potential}
	V = \Lambda^4\bqty{\frac{g \phi}{\Lambda}  + \pqty{\frac{g \phi}{\Lambda}}^2 + \ldots} - \Lambda^2 \pqty{\alpha - \frac{g \phi}{\Lambda}} \abs{H}^2
	+ \lambda\abs{H}^4 + \epsilon \Lambda_c^{4-n} v^n \cos(\frac{\phi}{f}) \text{,}
\end{equation}
where \(\phi\) is the relaxion, \(H\) is the Higgs doublet, \(g\) and \(\epsilon\) are small coupling constants, and \(\Lambda\), \(\Lambda_c\), \(v\) are respectively a~model's cutoff, an energy scale at which the cosine term originates and the electroweak scale. The remaining constants \(\alpha\) and \(\lambda\) are \(\order{1}\) parameters. It will become clear in the following part that the exact cosine form is not essential for the model to work, and a~wider class of functions could in principle be allowed here, provided that they produce appropriate extrema. The potential \eqref{eq:basic_potential} satisfies naturalness criteria, namely its parameters are \(\order{1}\) and there is no forced hierarchy between them. The small coupling constant \(g\) softly breaks the usual discrete shift symmetry \(\phi \to \phi + 2 \pi f \) of an axion, which would make \(\phi\) a~pseudo-Nambu-Goldstone boson. This symmetry is recovered in the limit of \(g \to 0\).

One could ask about a~situation in which the shift symmetry breaking in the Higgs' mass term is characterized by a~small coupling constant \(g_h\) different than \(g\). As explained in \cite{McAllister2016} in a~situation where \(g_h \gg g\) quantum corrections from Higgs loops would drive \(g\)~to a~value comparable with \(g_h\) anyway. On the other hand, if \(g \gg g_h\), then the natural excursion range would increase to \(\Lambda/g_h \gg \Lambda/g\). For an analysis of dynamical relaxation it is then sufficient to consider cases in which \(g \sim g_h\), a~class of solutions which requires the smallest excursion ranges.

The model requires no special choice of initial conditions, other than \(\phi\) being large enough to keep Higgs' mass-squared positive. Initially therefore the Higgs' vev \(v\) is zero and the electroweak symmetry is unbroken. As described in detail in \cite{Graham2015}  one can derive an estimate (up to a~chosen minimum) of the final electroweak scale
\begin{equation}
	v \simeq \Lambda_c \sqrt[n]{\frac{g \Lambda^3 f}{\epsilon \Lambda_c^4}} \text{.}
\end{equation}

\subsection{The double-scanner mechanism}

The double-scanner mechanism, presented in \cite{Espinosa2015}, extends the Lagrangian by adding a~second field \(\sigma\). The idea is that during the evolution Higgs' mass-squared will track \(\phi\), while \(\phi\) will track \(\sigma\). The potential, up to terms linear in \(\phi\) and \(\sigma\), is given by
\begin{subequations}
\begin{equation}\begin{split}
    \label{eq:v_double_scanner}
    V\pqty{\phi, \sigma, H} = {}
    &\Lambda^4 \pqty{\frac{g \phi}{\Lambda} + \frac{g_\sigma \sigma}{\Lambda}} - \Lambda^2 \pqty{\alpha - \frac{g \phi}{\Lambda}} \abs{H}^2 + \lambda \abs{H}^4 \\
    &+ A\pqty{\phi, \sigma, H} \cos(\frac{\phi}{f}) \text{,}
\end{split}\end{equation}
where
\begin{equation}
	\label{eq:v_double_scanner_ampl}
	A\pqty{\phi, \sigma, H} = \epsilon \Lambda^4 \pqty{\beta + c_\phi \frac{g \phi}{\Lambda} - c_\sigma \frac{g_\sigma\sigma}{\Lambda} + \frac{\abs{H}^2}{\Lambda^2}} \text{,}
\end{equation}
\end{subequations}
\(g\), \(g_\sigma\), \(\epsilon\) are small coupling constants and \(\alpha\), \(\beta\), \(c_\phi\), \(c_\sigma\) are \(\order{1}\) parameters. It is assumed that all terms are generated at a~cutoff scale \(\Lambda\).

\begin{figure}
\centering
\includegraphics[width=0.7\linewidth,trim=0 0 0 1.5cm, clip]{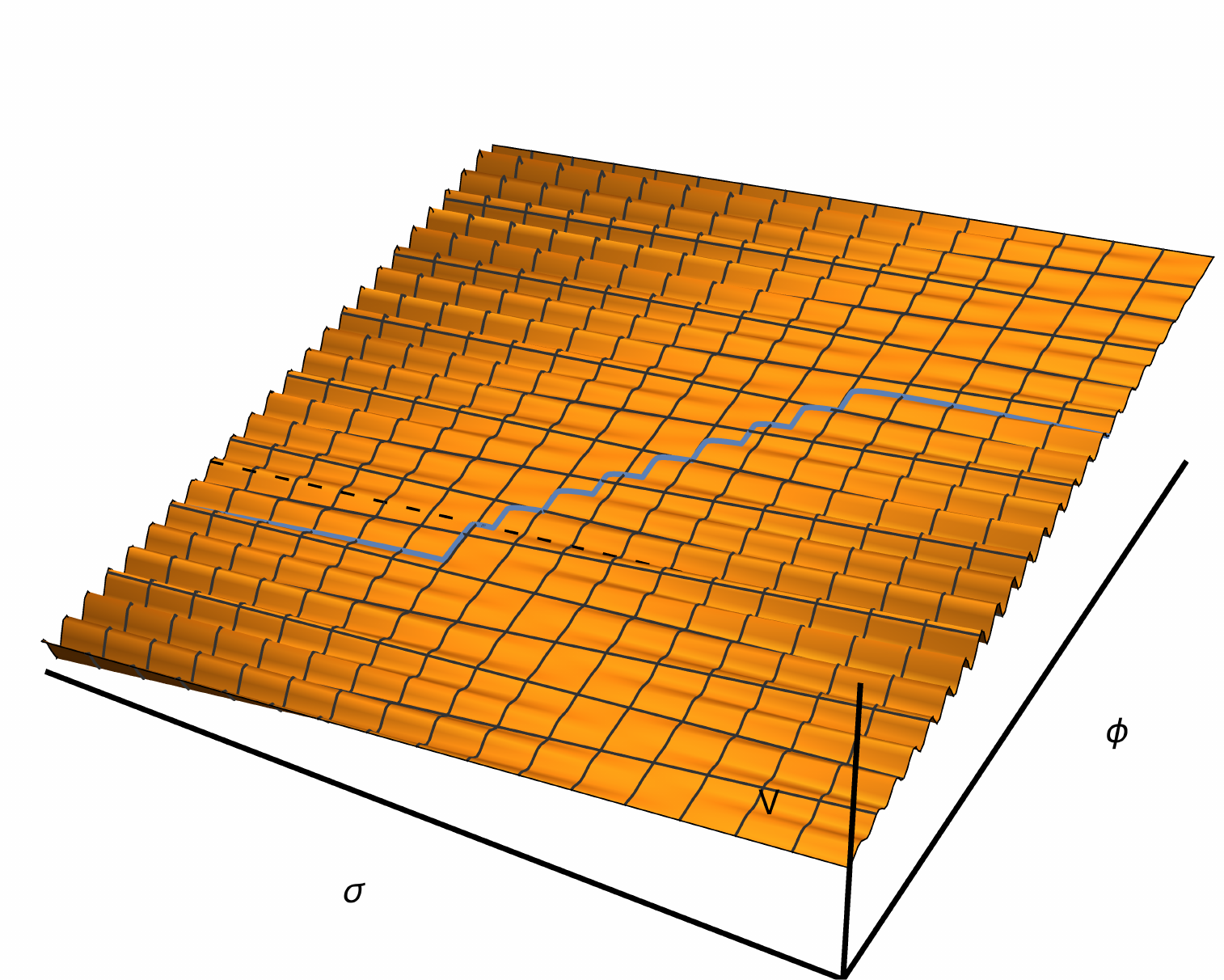}
\includegraphics[width=0.6\linewidth]{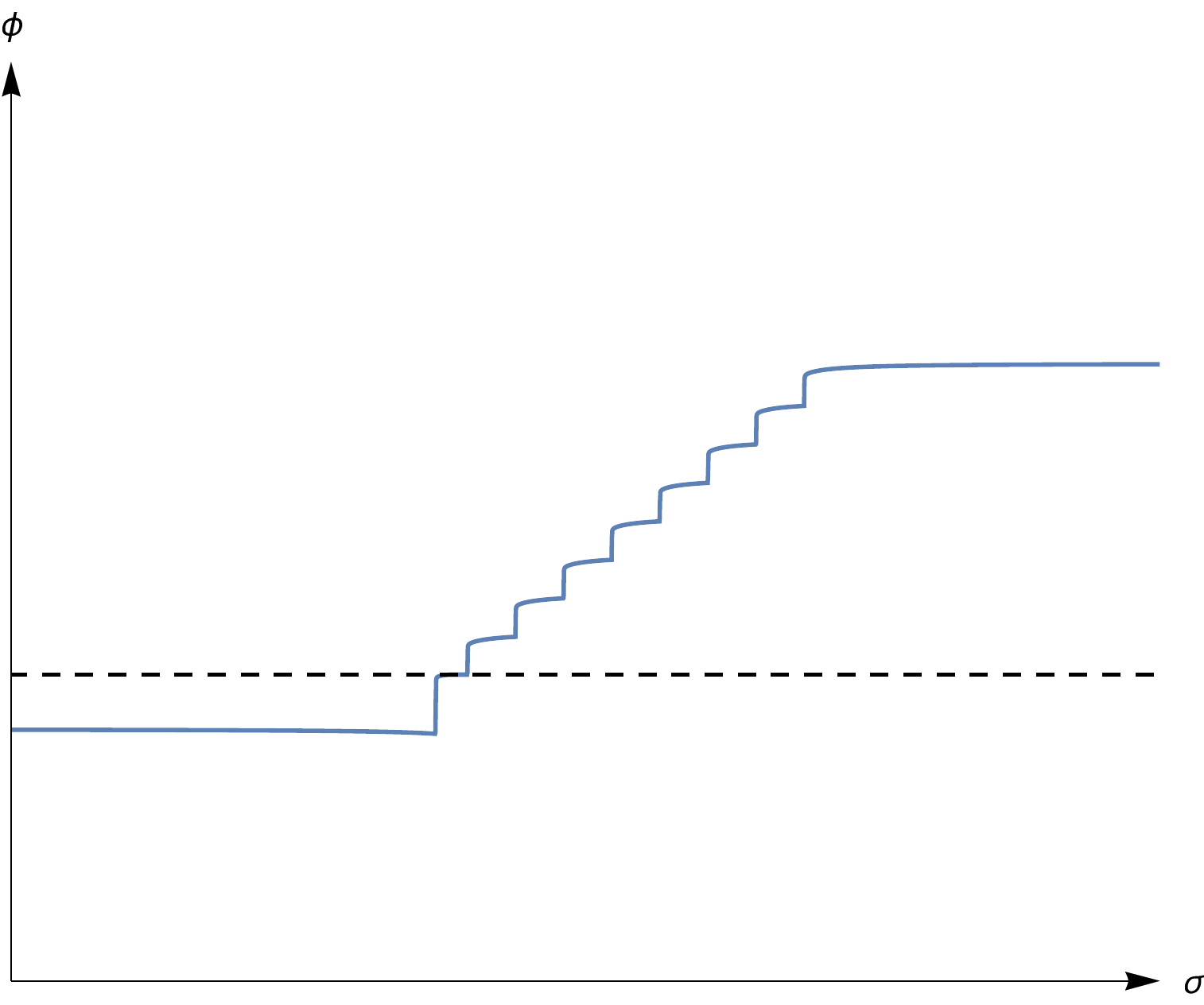}
\caption{The upper figure presents 3D picture of the potential of the double-scanner mechanism as a~function of the field \(\phi\) and \(\sigma\). One can see the characteristic periodic potential valleys responsible for controlling the evolution and a~flat band through which the fields slide at the center. The lower figure shows a~projection on the \(\pqty{\phi, \sigma}\) plane, giving a~clear view of the step-like behaviour. The dashed line indicates a critical value of \(\phi\) at which a~spontaneous breaking of the electroweak symmetry occurs. See also \cite{Espinosa2015}.}
\label{fig:evolution}
\end{figure}

\subsubsection{The no-minima band}

A plot of the potential is shown in Figure \ref{fig:evolution}. To analyze the vev selection process one needs the parameters of the no-minima band visible in the center, which is given by the condition
\begin{equation}
	\pdv{V}{\phi} > 0 \text{,}
\end{equation}
which, in the region where \(\expval{H} = 0\), leads to
\begin{equation}
	g \Lambda^3 \pqty{1 + \epsilon c_\phi \cos(\frac{\phi}{f})} > \frac{1}{f} A\pqty{\phi, \sigma, H} \sin(\frac{\phi}{f}) \text{.}
\end{equation}
As \(\epsilon \ll 1\) we can neglect the cosine term. Moreover we can get rid of the periodic dependence by taking a~modulus of both sides and approximating sine by its maximum value, which in turn produces a~simplified condition
\begin{equation}
	\label{eq:no_minima_condition}
	\abs{A\pqty{\phi,\sigma,H}} < g \Lambda^3 f \text{.}
\end{equation}
This relationship can be easily solved for \(\phi\)
\begin{equation}
	\label{eq:band_no_vev}
	\phi \in \phi_c + \frac{c_\sigma g_\sigma}{c_\phi g} \pqty{\sigma - \sigma_c} \pm \frac{f}{c_\phi \epsilon} \text{,}
\end{equation}
where \(\phi_c = \flatfrac{\Lambda}{g}\) is the critical value of \(\phi\) at which Higgs acquires its vev and \(\sigma_c = \flatfrac{\pqty{g c_\phi \phi_c + \beta \Lambda}}{\pqty{c_\sigma g_\sigma}}\). One can immediately see, that the band is described by a~simple linear function, with a~width proportional to the cosine's period~\(f\).

When the Higgs' vev is nonzero, one has to consider terms proportional to \(\abs{H}^2\) and \(\abs{H}^4\). In this case the condition gives
\begin{equation}
	g\Lambda^3 + \frac{g\Lambda^3}{2\lambda}\pqty{\alpha - \frac{g\phi}{\Lambda}} + \pdv{A}{\phi}\cos(\frac{\phi}{f}) > \frac{1}{f} A\pqty{\phi,\sigma,H}\sin(\frac{\phi}{f}) \text{.}
\end{equation}
As we are interested in a~behaviour near the critical point \(\phi = \phi_c\) we can express the second term by
\begin{equation}
	\frac{g^2 \Lambda^2}{2\lambda} \, \Delta\phi \text{,}
\end{equation}
where \(\Delta\phi\) is given by \(\phi_c - \phi\). From this one can see, that it is of order of \(g^2\) and in the limit \(g \ll 1\) can be neglected. We therefore reproduce the condition \eqref{eq:no_minima_condition}. Again, solving for \(\phi\) one obtains
\begin{equation}
	\label{eq:band_with_vev}
	\phi \in \phi_c + \frac{c_\sigma g_\sigma}{c'_\phi g} \pqty{\sigma - \sigma_c} \pm \frac{f}{c'_\phi \epsilon} \text{,}
\end{equation}
where \(c'_\phi = c_\phi - \flatfrac{1}{\pqty{2\lambda}}\). This solution obviously continuously merges with \eqref{eq:band_no_vev} on the critical line, where the Higgs boson acquires its vev.

\subsubsection{Evolution stages}
\label{sec:evolution_stages}

\paragraph{Stage I~(\(\phi = \text{const}\), \(\expval{H} = 0\))}

\noindent Initially values of both \(\phi\) and \(\sigma\) are large, such that the mass-squared of the Higgs field is positive and the cosine amplitude is negative. The field \(\sigma\) now slow-rolls just as \(\phi\) did in the original model, with a~friction provided by the continuing inflation
\begin{equation}
	\label{eq:sigma_evolution}
	\sigma(t) = - \frac{g_\sigma \Lambda^3}{3 H_I}t + C_1 e^{-3 H_I t} + \text{const} \text{.}
\end{equation}
Once again the exponential part is assumed to vanish quickly, removing a~possible dependence on the initial conditions. At the same time \(\phi\) is stuck in one of the minima produced by the periodic term. This situation persists until \(\sigma\) reaches the no-minima band, as can be seen on Figure \ref{fig:evolution}, and the amplitude of the cosine term becomes too small to hold \(\phi\) any longer.

\paragraph{Stage II (\(\phi \neq \text{const}\), \(\expval{H} = 0\))}

\noindent When \(\phi\) drops out of a~minimum, it proceeds to roll down according to
\begin{equation}
	\phi\pqty{t} \simeq -\frac{g\Lambda^3}{3H_I}t + \text{const} \text{,}
\end{equation}
where small effects of the periodic term which are responsible for the step-like behavior (which are \(\order{f}\)) have been neglected. After some time \(\phi\) encounters another potential hill and stabilizes again, where the necessary condition is that the evolution's trajectory gradient in the \(\pqty{\phi,\sigma}\) plane is larger than the gradient of the no-minima band \eqref{eq:band_no_vev}
\begin{equation}
	\frac{\dv*{\phi}{t}}{\dv*{\sigma}{t}} > \frac{c_\sigma g_\sigma}{c_\phi g} \text{,}
\end{equation}
which leads to a~condition
\begin{equation}
	\label{eq:couplings_relationship}
	c_\phi g^2 > c_\sigma g_\sigma^2 \text{.}
\end{equation}
The process repeats step-by-step (as visible in Figure \ref{fig:evolution}.), with \(\phi\) slowly decreasing, effectively tracking the evolution of \(\sigma\).

\paragraph{Stage III (\(\phi \neq \text{const}\), \(\expval{H} > 0\))}

\noindent When \(\phi\) crosses the critical value \(\phi_c = \flatfrac{\alpha\Lambda}{g}\), the mass-squared of the Higgs field becomes negative and a~spontaneous breaking of the electroweak symmetry occurs. The (now nonzero) Higgs vev contributes to the overall amplitude of the cosine, changing the slope of the no-minima band, as shown in \eqref{eq:band_with_vev}. \(\phi\) and \(\sigma\) now slide across the band until its another edge is reached, continuously increasing the vev. For the trajectory to exit the band on the other side it is required that
\begin{equation}
	\label{eq:couplings_relationship_2}
	c'_\phi g^2 < c_\sigma g_\sigma^2 \text{.}
\end{equation}

\paragraph{Stage IV (\(\phi = \text{const}\), \(\expval{H} > 0\))}

\noindent Finally, the trajectory exits the no-minima band, and \(\phi\) again enters one of the minima produced by the periodic term. As \(\phi\)'s value is now fixed, so is the electroweak scale. The selection process is thus completed. The field \(\sigma\) continues to roll down until it reaches its own minimum.

\subsubsection{The final electroweak scale}

As potential valleys of \(\phi\) are narrow with respect to the whole field range traversed, one can in fact assume that they are infinitely dense. In this approximation the possible final values of \(\phi\) are continuous (the evolution trajectory can exit the no-minima band at any place) and it is possible to find an explicit formula for the final electroweak scale in terms of the model's parameters.

\begin{figure}
\centering
\begin{tikzpicture}[scale=0.5]
    \fill[fill=green,opacity=0.2] (1,6) -- (-2,0) -- (1,-6) -- (5,-6) -- (2,0) -- (5,6);
    \draw (1,6) -- (-2,0) -- (1,-6);
    \draw[dashed] (3,6) -- (0,0) node[below left] {$(\sigma_0,\phi_0)$} -- (3,-6);
    \filldraw (0,0) circle (2pt);
    \draw (5,6) -- (2,0) -- (5,-6);
    
    \draw[dashed] (-6,0) node[left] {$\phi_c$} -- (8,0);
    \draw (-1,0) node[above] {$x$} (1,0) node[above] {$x$};
    
    \draw[-latex,blue,thick] (5,6) -- (3.5,3);
    \draw[-latex,blue,thick] (5,6) -- (2,0) -- (1.25,-2.5);
    \draw[-latex,blue,thick] (2,0) -- (0.5,-5) -- (-3,-5);
    \draw[blue,thick] (0.5,-5) -- (-6,-5) node[left,black] {$\phi_f$};
    
    \draw (7,3) node {I} (7,-3) node {II};
\end{tikzpicture}
\caption{Approximate evolution in the \(\pqty{\sigma, \phi}\) plane (blue line). Evolution starts in the region I~where \(v\) is 0. When \(\phi\) crosses the critical value \(\phi_c\), the Higgs doublet acquires vev and a~spontaneous symmetry breaking occurs. As \(\phi\)~goes down in region II \(v\) continues to increase. Finally \(\phi\) exits the band settling down with a final value \(\phi_f\) and hence completing the vev selection process.}
\label{fig:final_vev}
\end{figure}

The relevant region of the field space \(\pqty{\phi,\sigma}\) in presented in Figure \ref{fig:final_vev}. As the trajectory and the band are described by linear functions with known gradients, finding the final value \(\phi_f\) of \(\phi\) amounts to a~simple geometrical task of locating their intersection. The fixed points are \(\pqty{\sigma_c - x, \phi_c}\) and \(\pqty{\sigma_c + x, \phi_c}\), where \(x = \flatfrac{gf}{\pqty{\epsilon c_\sigma g_\sigma}}\), and the respective gradients are \(\flatfrac{g}{g_\sigma}\) and \(\flatfrac{c_\sigma g_\sigma}{\pqty{c'_\phi g}}\). From this data one can find that the \(\phi\) coordinate of the intersection (the exit point) is given by
\begin{equation}
	\phi_f = \frac{\alpha\Lambda}{g} - \frac{2fg^2}{\epsilon\pqty{c_\sigma g_\sigma^2 - c'_\phi g^2}} + \order {f}\text{.}
\end{equation}
The electroweak scale can be obtained by putting the above result to the formula for the vev:
\begin{equation}\begin{split}
	\label{eq:final_vev}
	v^2 
	&= - \frac{\Lambda^2}{\lambda}\pqty{\alpha - \frac{g\phi}{\Lambda}} \\
	&= \frac{4 g^3 \Lambda f}{\epsilon\lambda\pqty{c_\sigma g_\sigma^2 - c_\phi g^2 + \frac{g^2}{2\lambda}}} + \order{g\Lambda f} \text{.}
\end{split}\end{equation}
From \eqref{eq:final_vev} one can explicitly see that the model produces an electroweak scale that is proportional to the coupling~\(g\). It can therefore naturally result in a~scale much smaller than the cutoff, provided that the coupling constant is small enough. It is also important to note that not all parameters of the model contribute to the selected vev, as \(\alpha\) and \(\beta\) are not present in the final formula.

\subsubsection{Consistency requirements}

As shown in \cite{Espinosa2015} several constraints must be fulfilled for the model to work consistently. These constraints do apply to the multi-Higgs models discussed in this paper. In particular,  relaxation imposes severe requirements on the inflation sector. The required inflation has to be both slow (to ensure that classical evolution is dominant) and very long (to give relaxation enough time to complete). It is than a~question whether one can build an inflation model that satisfies those requirements without excessive fine-tuning which relaxation is designed to avoid. For existing attempts in this field see \cite{DiChiara2015,Patil2015,Marzola2015,Fowlie2016}. An~alternative, where the necessary friction is provided by particle production without a~need of inflation see \cite{Hook2016}. These mechanisms can easily be extended to the multi-Higgs models discussed here.

\section{Relaxation in 2HDM}

Generic potential for a~relaxion mechanism in 2HDM has the form
\begin{equation}
	V = V\pqty{\phi, \sigma} + V_{2HDM}\pqty{\phi, \Phi_1, \Phi_2} + A\pqty{\phi, \sigma, v}\cos(\frac{\phi}{f}) \text{.}
\end{equation}
As in the single doublet model the first part describes the slow rolling of the scalar fields \(\phi\) and \(\sigma\) when vevs of both Higgs doublets vanish and the amplitude of the periodic term is small. It is assumed to be the same as in the single doublet scenario, therefore the analysis presented in section \ref{sec:evolution_stages} for the most part holds here as well. The second term is the ordinary potential of 2HDM supplemented by a~coupling to \(\phi\), which makes vevs depend on its value. Finally there is a~periodic term, whose amplitude depends on the scalar fields  and on doublets' vevs.

\subsection{2HDM potential}

The general potential for a~multi-Higgs doublet model, where all doublets have the same transformation properties, has the form:
\begin{equation}
	V = Y_{ab} \Phi_a^\dagger \Phi_b + Z_{abcd} \Phi_a^\dagger \Phi_b \Phi_c^\dagger\Phi_d \text{,}
\end{equation}
with symmetry conditions
\begin{subequations}\begin{gather}
	Y_{ab} = Y_{ba}^* \text{,} \\
	Z_{abcd} = Z_{cdab} \text{,} \\
	Z_{abcd} = Z_{badc}^* \text{.}
\end{gather}\end{subequations}
For two doublets (2HDM) the potential can be presented as:
\begin{equation}\begin{split}
	\label{eq:potential}
	V_{2HDM} = {}
	& m^2_{11} \Phi_1^\dagger \Phi_1 + m^2_{22} \Phi_2^\dagger \Phi_2 - \pqty{m^2_{12} \Phi_1^\dagger \Phi_2 + \text{h.c.}} \\
	&+ \frac{\lambda_1}{2}\pqty{\Phi_1^\dagger \Phi_1}^2 + \frac{\lambda_2}{2}\pqty{\Phi_2^\dagger \Phi_2}^2 + \lambda_3\pqty{\Phi_1^\dagger \Phi_1}\pqty{\Phi_2^\dagger \Phi_2} + \lambda_4\pqty{\Phi_1^\dagger \Phi_2}\pqty{\Phi_2^\dagger \Phi_1} \\
	& + \bigg[ \frac{\lambda_5}{2}\pqty{\Phi_1^\dagger \Phi_2}^2 + \lambda_6\pqty{\Phi_1^\dagger \Phi_1}\pqty{\Phi_1^\dagger \Phi_2} + \lambda_7\pqty{\Phi_2^\dagger \Phi_2}\pqty{\Phi_1^\dagger \Phi_2} + \text{h.c.} \bigg] \text{,}
\end{split}\end{equation}
where parameters \(m^2_{11}\), \(m^2_{22}\), \(\lambda_1\), \ldots, \(\lambda_4\) are real and \(m^2_{12}\), \(\lambda_5\), \ldots, \(\lambda_7\) are complex. By selecting an appropriate isospin axis and using the overall phase freedom one can, without loss of generality, choose the following representation for the doublets' vevs~\cite{Ginzburg2005}:
\begin{equation}
	\expval{\Phi_1} = \pmqty{0 \\ v_1} \text{,} \hspace{1cm}
	\expval{\Phi_2} = \pmqty{u \\ v_2 e^{i\xi}} \text{,}
\end{equation}
with \(v_1,v_2\) being real. Furthermore as charge breaking vacuum would be unphysical one can take \(u = 0\). This does not exhaust the freedom one has in the choice of the Lagrangian form. With both Higgs doublets having the same quantum numbers a~global unitary transformation can be performed
\begin{equation}
	\pmqty{\Phi_1' \\ \Phi_2'} = \mathcal{F} \pmqty{\Phi_1 \\ \Phi_2} \text{,}
\end{equation}
where
\begin{equation}
	\label{eq:basis_transformation}
	\mathcal{F} = e^{-i\rho_0} \pmqty{\cos\theta e^{i\rho/2} & \sin\theta e^{i(\tau - \rho/2)} \\
	- \sin\theta e^{-i(\tau - \rho/2)} & \cos\theta e^{-i\rho/2}} \text{.}
\end{equation}
Such transformation can be used, by choosing \(\rho = \xi\), to remove the complex phase from the second vev, hence producing a~``real vacuum'' basis \cite{Ginzburg2005}.

For the dynamic relaxation model an additional dependence on \(\phi\) is introduced in the quadratic term:
\begin{equation}
	\label{eq:mass_squared}
	m^2_{ab} = -\Lambda^2\pqty{\alpha_{ab} - \gamma_{ab}\frac{g\phi}{\Lambda}} \text{,}
\end{equation}
where \(\alpha_{ab}\) and \(\gamma_{ab}\) are arbitrary hermitian matrices with \(\order{1}\) entries. The spontaneous symmetry breaking occurs when at least one eigenvalue of \(m_2\) is negative, which will happen when \(\phi\) achieves a~low-enough value. It is useful to use the SM constraint \(v^2 = v_1^2 + v_2^2\) to express vevs as:
\begin{equation}\begin{split}
	\label{eq:vev_representation}
	v_1 &= v \cos\beta\text{,} \\
	v_2 &= v \sin\beta \text{.}
\end{split}\end{equation}
The potential minimum conditions then take the form \cite{Davidson2005}:
\begin{subequations}
\begin{equation}\begin{split}
	\label{eq:minimum_conditions}
	m^2_{11} = m^2_{12}\tan\beta - \frac{v^2}{2}\Big[
	&\lambda_1\cos^2\beta + \pqty{\lambda_3 + \lambda_4 + \lambda_5}\sin^2\beta \\
    {} &+ \pqty{2\lambda_6 + \lambda_6^*}\sin\beta\cos\beta + \lambda_7\sin^2\beta\tan\beta \Big] \text{,}
\end{split}\end{equation}
\begin{equation}\begin{split}
	m^2_{22} = \pqty{m^2_{12}}^*\tan^{-1}\beta - \frac{v^2}{2}\Big[
	&\lambda_2\sin^2\beta + \pqty{\lambda_3 + \lambda_4 + \lambda_5^*}\cos^2\beta \\
    &+ \lambda_6^*\cos^2\beta\tan^{-1}\beta + \pqty{\lambda_7 + 2\lambda_7^*}\sin\beta\cos\beta\Big] \text{.}
\end{split}\end{equation}
\end{subequations} 

One also needs to consider the dependence of the cosine's amplitude on the vevs. In the most general 2HDM scenario it is also described by an \(\order{1}\) hermitian matrix. This gives the amplitude a~from
\begin{equation}
	A\pqty{\phi,\sigma,\Phi_1,\Phi_2} = \Lambda^4 \epsilon \pqty{\beta + c_\phi \frac{g\phi}{\Lambda} - c_\sigma \frac{g_\sigma\sigma}{\Lambda} + \frac{\rho_{ab}}{\Lambda^2} \abs{\Phi_a} \abs{\Phi_b}} \text{.}
\end{equation}
The three matrices \(\alpha_{ab}\), \(\gamma_{ab}\) and \(\rho_{ab}\) are not uniquely defined. Elements of \(\rho_{ab}\) can be rescaled by absorbing a~factor into \(\epsilon\). What is even more important, we can use transformations \eqref{eq:basis_transformation} to adjust the choice of basis in such a~way that one of these matrices is diagonal.

In the usual 2HDM one can make such choice of the angle \(\beta\), that one of the vevs is set to zero (so called Higgs basis). In this case only a~single electroweak scale \(v\) is left, which greatly simplifies computations. Unfortunately, for the dynamic relaxation this route cannot be used. The angle \(\beta\) would be explicitly present in the amplitude. Although theoretically given by \eqref{eq:minimum_conditions}, in general the analytic solution cannot be obtained. One could try diagonalizing the mass-squared matrix in the hope that it would give vevs in a~non-Higgs basis. In this case however the diagonalization would depend on \(\phi\) and hence on time as well. This time dependence would propagate to the coupling parameters, making direct results practically impossible.

However, it is possible to find the constraints under which the electroweak symmetry breaking happens at all. The critical points in the cosmological evolution correspond to the situation when
\begin{equation}
	\det(m^2) = \det(\alpha_{ab} - \gamma_{ab}\frac{g\phi}{\Lambda}) = 0 \text{.}
\end{equation}
This would result in a~quadratic equation for the critical values \(\phi_c\). For existence of real solution it is required that
\begin{equation}
	\pqty{\alpha_{22} \gamma_{11} + \alpha_{11}\gamma_{22} - 2\Re{\alpha_{12}\gamma_{12}}}^2 - 4\pqty{\abs{\alpha_{12}}^2 - \alpha_{11}\alpha_{22}} \pqty{\abs{\gamma_{12}}^2-\gamma_{11}\gamma_{22}} \ge 0 \text{.}
\end{equation}
In the special case of equality only one vev would be generated\footnote{The possibility of only one nonzero vev remains valid even if the value is strictly greater than~0. It is possible for \(\phi\) to exit the band before the second vev is generated.}.

\subsection{2HDM with additional symmetries}

In a~situation when additional symmetries are enforced it is possible to completely solve the dependence of the vacuum expectation values of the Higgs fields on parameters of the model. What is required is that in the potential there are no terms with odd powers of a~Higgs doublet. This could be accomplished by giving the doublets different U(1) charges, making one of them odd under an internal $Z_2$ parity transformation or by putting scalar fields into different SU(2) representations. In this case the potential simplifies to:
\begin{equation}\begin{split}
	V(\phi, \sigma, H_1, H_2)
	&= \Lambda^4\pqty{\frac{g\phi}{\Lambda} + \frac{g_\sigma\sigma}{\Lambda}} \\
	&- \Lambda^2 \pqty{\alpha_1 - \frac{g\phi}{\Lambda}}\abs{H_1}^2  + \lambda_1 \abs{H_1}^4 \\
	&- \Lambda^2 \pqty{\alpha_2 - \frac{g\phi}{\Lambda}}\abs{H_2}^2  + \lambda_2 \abs{H_2}^4 \\
	&+ \lambda_3 \abs{H_1}^2 \abs{H_2}^2 \\
	&+ A\pqty{\phi, \sigma, H_1, H_2} \cos(\frac{\phi}{f}) \text{,}
\end{split}\end{equation}
with the amplitude given by
\begin{equation}
    \label{eq:2hdm_amplitude}
	A\pqty{\phi, \sigma, H_1, H_2} = \epsilon\Lambda^4 \pqty{\beta + c_\phi\frac{g\phi}{\Lambda} - c_\sigma\frac{g_\sigma\sigma}{\Lambda} + c_1\frac{\abs{H_1}^2}{\Lambda^2} + c_2\frac{\abs{H_2}^2}{\Lambda^2}} \text{.}
\end{equation}
As vevs of the two doublets could enter the amplitude with different strengths the formula was supplemented with additional \(\order{1}\) constants \(c_1\), \(c_2\). One could always absorb one of these constants into \(\epsilon\), here however they are kept to easily track their influence on the final result. A possible issue with the potential is that it contains flat directions, which result in physically unacceptable massless scalars. For the following discusison we assume that those directions are frozen, or that the masses are generated by one of the mechanisms described in \cite{Dev2014}.

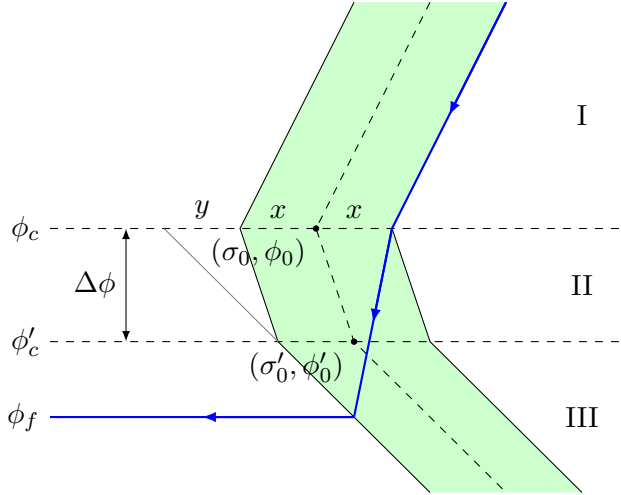
\begin{figure}
\centering
\begin{tikzpicture}[scale=0.5]
    \fill[fill=green,opacity=0.2] (1,6) -- (-2,0) -- (-1,-3) -- (3,-7) -- (7,-7) -- (3,-3) -- (2,0) -- (5,6);
    \draw (1,6) -- (-2,0) -- (-1,-3) -- (3,-7);
    \draw[dashed] (3,6) -- (0,0) node[below left] {\(\pqty{\sigma_0,\phi_0}\)} -- (1,-3) node[below left] {\(\pqty{\sigma_0',\phi_0'}\)} -- (5,-7);
    \filldraw (0,0) circle (2pt) (1,-3) circle (2pt);
    \draw (5,6) -- (2,0) -- (3,-3) -- (7,-7);
    
    \draw[dashed] (-7,0) node[left] {\(\phi_c\)} -- (8,0);
    \draw (-1,0) node[above] {\(x\)} (1,0) node[above] {\(x\)};
    \draw[dashed] (-7,-3) node[left] {\(\phi_c'\)} -- (8, -3);
    \draw[latex-latex] (-5,0) -- (-5,-3);
    \draw (-5,-1.5) node[left] {$\Delta\phi$};
    
    \draw[-latex,blue,thick] (5,6) -- (3.5,3);
    \draw[-latex,blue,thick] (5,6) -- (2,0) -- (1.5,-2.5);
    \draw[-latex,blue,thick] (2,0) -- (1,-5) -- (-3,-5);
    \draw[blue,thick] (1,-5) -- (-7,-5) node[left,black] {\(\phi_f\)};
    
    \draw[gray] (-4,0) -- (-1,-3);
    \draw (-3,0) node[above] {\(y\)};
    
    \draw (7,3) node {I} (7,-1.5) node {II} (7,-5) node {III};
\end{tikzpicture}
\caption{Approximate evolution in the plane \(\pqty{\sigma,\phi}\). The green band is the region where \(\phi\) has no minima. Evolution starts in region I, where vevs of both Higgs doublets are 0. When \(\phi\) crosses \(\phi_c\), \(H_1\) starts obtaining vev, and through backreaction changes the band slope. The same happens with \(H_2\) after \(\phi\) crosses \(\phi_c'\). Ultimately \(\phi\) exits the band and stabilizes at some minimum with a~value \(\phi_f\) also fixing vevs of both doublets.}
\label{fig:evolution_2hdm}
\end{figure}

The Higgs fields acquire vevs through spontaneous symmetry breaking when their respective mass-squared terms become negative. For \(H_1\) it takes place when \(\phi\) crosses a~critical value \(\phi_c\) and for \(H_2\) when \(\phi\) crosses \(\phi_c'\). After SSB the vevs are given by:
\begin{subequations}\begin{gather}
	\abs{H_1}^2 = \frac{\Lambda^2}{2\lambda_1}\pqty{\alpha_1 - \frac{g\phi}{\Lambda} - \lambda_3\frac{\abs{H_2}^2}{\Lambda^2}} \text{,} \\
	\abs{H_2}^2 = \frac{\Lambda^2}{2\lambda_2}\pqty{\alpha_2 - \frac{g\phi}{\Lambda} - \lambda_3\frac{\abs{H_1}^2}{\Lambda^2}} \text{.}
\end{gather}\end{subequations}
The vevs contribute to one another, therefore to make use of these formulae one must first decouple the dependence, which leads to:
\begin{subequations}
\begin{gather}
	\label{eq:2hdm_vev_from_phi_1}
	\abs{H_1}^2 = \frac{\Lambda^2}{2\lambda_1}\pqty{1 - \frac{\lambda_3^2}{4\lambda_1\lambda_2}}^{-1} \pqty{\alpha_1 - \frac{\lambda_3}{2\lambda_2}\alpha_2 - \pqty{1 - \frac{\lambda_3}{2\lambda_2}} \frac{g\phi}{\Lambda}} \text{,} \\
	\label{eq:2hdm_vev_from_phi_2}
	\abs{H_2}^2 = \frac{\Lambda^2}{2\lambda_2} \pqty{1 - \frac{\lambda_3^2}{4\lambda_1\lambda_2}}^{-1} \pqty{\alpha_2 - \frac{\lambda_3}{2\lambda_1}\alpha_1 - \pqty{1 - \frac{\lambda_3}{2\lambda_1}} \frac{g\phi}{\Lambda}} \text{.}
\end{gather}
\end{subequations}
As in the model with one doublet the task of finding the final vevs amounts to determining the field-space coordinates at which the evolution trajectory exits the no-minima band. The situation was presented on figure \ref{fig:evolution_2hdm}, where it is assumed that \(H_2\) is the lighter doublet (the one that acquires a~vev later).

The critical values of \(\phi\) can be easily found from \eqref{eq:2hdm_vev_from_phi_1} and \eqref{eq:2hdm_vev_from_phi_2}. They are respectively:
\begin{subequations}\begin{gather}
	\phi_c = \frac{\Lambda}{g} \alpha_1 \text{,} \\
	\phi_c' = \frac{\Lambda}{g} \pqty{\alpha_2 - \frac{\lambda_3 \Delta\alpha}{2\lambda_1 - \lambda_3}} \text{,}
\end{gather}\end{subequations}
where \(\Delta\alpha = \alpha_1 - \alpha_2\). For two doublets to obtain a vev it is necessary that 
both critical values are crossed before the relaxation process is completed. This will happen if
\begin{equation}
\frac{2fg^2}{\epsilon\pqty{c_\sigma g_\sigma^2 - c_\phi^{II} g^2}} < \frac{\Lambda}{g} \pqty{1 + \frac{\lambda_3}{2\lambda_1 - \lambda_3}} \Delta\alpha \text{.}
\end{equation}
If one sets all \(\order{1}\) parameters to 1, \(\Lambda \sim f\), \(g \sim g_\sigma\), a convenient order-of-magnitude criterion is obtained:
\begin{equation}\label{eq:two_vevs}
\frac{g}{\epsilon} \sim \frac{v^2}{\Lambda^2} \gtrsim \Delta\alpha \text{,}
\end{equation}
which puts severe constraint on the difference between parameters \(\alpha_i\). If the above condition is not satisfied, then the first doublet to obtain a vev would stop the relaxation, leaving the other one vevless, with a mass of the order of \(\Lambda\). It is worth noting, that although in this situation addition of the second doublet does not introduce significant differences with respect to the single Higgs relaxation, it naturally produces 2HDM with a small electroweak scale, with one doublet being massive and inert.

The following analysis assumes that the condition \eqref{eq:two_vevs} is satisfied. A key ingredient is the band gradient in regions II and III, which is given by
\begin{equation}
	\dv{\phi^*}{\sigma} = \begin{cases}
	\frac{c_\sigma g_\sigma}{c_\phi^{II}}   & \phi_c' < \phi < \phi_c \\
	\frac{c_\sigma g_\sigma}{c_\phi^{III}}  & \phi < \phi_c'
	\end{cases} \text{,}
\end{equation}
where \(c_\phi^{II} = c_\phi - \flatfrac{1}{\pqty{2\lambda_1}}\) and \(c_\phi^{III} = c_\phi - \flatfrac{1}{\pqty{2\lambda_1}} - \flatfrac{1}{\pqty{2\lambda_2}}\). Similarly to \eqref{eq:couplings_relationship_2} in the one doublet case for the trajectory to exit the band it is required that
\begin{equation}
	c_\phi^{III} g^2 < c_\sigma g_\sigma^2 \text{.}
\end{equation}
As both \(\lambda_1\) and \(\lambda_2\) enter here this produces weaker requirements on the parameters. The \(\phi\) coordinate of the exit point is
\begin{equation}
	\label{eq:2hdm_phi_f}
	\phi_f = \phi_c - \frac{2x - y}{\frac{g_\sigma}{g} - \frac{c_\phi^{III} g}{c_\sigma g_\sigma}} \text{,}
\end{equation}
where \(x = \flatfrac{gf}{\pqty{\epsilon c_\sigma g_\sigma}}\). The value of \(y\) can be found from the geometrical condition
\begin{equation}
	y = \pqty{\frac{c_\phi^{II} g}{c_\sigma g_\sigma} - \frac{c_\phi^{III} g}{c_\sigma g_\sigma}} \Delta\phi \text{,}
\end{equation}
and it is given by
\begin{equation}
	\label{eq:2hdm_y}
	y = \frac{c_2\Lambda}{c_\sigma g_\sigma} \Delta\alpha \frac{2\lambda_1}{4\lambda_1\lambda_2 - \lambda_3^2} \text{.}
\end{equation}
By combining \eqref{eq:2hdm_vev_from_phi_1}, \eqref{eq:2hdm_vev_from_phi_2}, \eqref{eq:2hdm_phi_f} and \eqref{eq:2hdm_y} we finally obtain:
\begin{subequations}
\begin{gather}
	v_1^2 = \frac{\Lambda^2}{\lambda_1}\pqty{1 - \frac{\lambda_3^2}{4\lambda_1\lambda_2}}^{-1} \Bigg[ \frac{\lambda_3}{2\lambda_2}\Delta\alpha + \pqty{1 - \frac{\lambda_3}{2\lambda_2}} \frac{\frac{2 g^3 f}{\epsilon\Lambda} - c_2 g^2 \frac{2\lambda_1}{4\lambda_1\lambda_2 - \lambda_3^2}\Delta\alpha}{c_\sigma g_\sigma^2 - c_\phi^{III} g^2} \Bigg] \text{,} \\
	v_2^2 = \frac{\Lambda^2}{\lambda_2}\pqty{1 - \frac{\lambda_3^2}{4\lambda_1\lambda_2}}^{-1} \Bigg[ -\Delta\alpha + \pqty{1 - \frac{\lambda_3}{2\lambda_1}} \frac{\frac{2 g^3 f}{\epsilon\Lambda} - c_2 g^2 \frac{2\lambda_1}{4\lambda_1\lambda_2 - \lambda_3^2}\Delta\alpha}{c_\sigma g_\sigma^2 - c_\phi^{III} g^2} \Bigg] \text{.}
\end{gather}
\end{subequations}
The above formulae show, that in the analyzed 2HDM scenario the final vevs of the doublets contain also terms proportional to \(\Delta\alpha\) that are not explicitly supressed by the small coupling \(g\). However, taking into account the condition \eqref{eq:two_vevs} those terms must be small as well.

The smallness of \(\Delta\alpha\) can be naturally explained in the context of symmetry-constrained 2HDMs \cite{Ivanov2007,Branco2012}. Out of the six Ivanov's symmetry classes three enforce the required cancellation: U(2) rotations in the doublet space and two kinds of the generalized CP symmetries. Unfortunately, constructing physically viable models involving such symmetries is difficult, especially once the quark sector is considered.

Given that of particular interest would be the maximally symmetric 2HDM (MS-2HDM) \cite{Dev2014}. In this model the Higgs sector potential
\begin{equation}
V = - m^2\pqty{\phi}\pqty{\abs{H_1}^2 + \abs{H_2}^2} + \lambda\pqty{\abs{H_1}^2 + \abs{H_2}^2}^2
\end{equation}
is characterized by only two parameters \(m^2\) and \(\lambda\), leading to \(\Delta\alpha = 0\), which makes it suitable for the dynamical relaxation. Small misalignment is produced once the model's symmetry is broken by the RGE running and by a soft mass term \(m_{12}\abs{H_1}\abs{H_2}\), which ensures that no massless states are present in the low energy limit (see the remarks below the formula \eqref{eq:2hdm_amplitude}).

\section{Conclusions}

The dynamical relaxation provides an interesting solution to the hierarchy problem in face of the missing signatures of any new physics in recent experiments. Through a~dynamical process taking place in the inflatory phase of the universe it manages to achieve a~small electroweak scale without introducing new states observable in~current experiments. As such it offers a~way to maintain naturalness and remain in agreement with experimental data.

Appropriate approximations made it possible to derive an explicit formula for the final Higgs' vev in the double-scanning scenario. It confirms the earlier result that the hierarchy between the electroweak scale and the model's cutoff originates from the smallness of the coupling constant softly-breaking the relaxion's shift symmetry. Moreover it is interesting to see that the chosen scale is independent of some of the parameters present in the Lagrangian, effectively allowing them to take any value without affecting the observational picture.

The analysis of the relaxation in 2HDM indicates, that in a~general case the relaxation would be stopped by the first doublet that gains a vev, with the other one remaining vevless with a mass of the order of \(\Lambda\). This happens to conform with the inert doublet model. The second doublet becomes notrivially involved in the symmetry constrained variants of 2HDM, whenever quantum corrections to both doublets at the cutoff are almost exactly equal, with MS-2HDM being an example that particularly well fits the relaxation mechanism.

It is important to point out that although dynamical relaxation is considered as a~mechanism for the selection of the electroweak scale, the whole mechanism is more general and can in principle be used to drive evolution of a~vev of any scalar field. In particular Lagrangians used here depend eventually only on the modulus of the Higgs doublet expectation value, which allows to exchange it for a~singlet without any change in the final formulae derived.

\vspace{\baselineskip}
\noindent {\bf Acknowledgements}\\
This work has been supported by National Science Centre under research grant DEC-2012/04/A/ST2/00099.

\end{document}